\documentclass[aps,prl,twocolumn,superscriptaddress]{revtex4}
\usepackage{graphics}
\usepackage{epstopdf}
\usepackage{graphicx}
\usepackage{dcolumn}
\usepackage{bm}
\usepackage{amsmath}
\usepackage{color}
\usepackage{ulem}

\newcommand{\be}{\begin{equation}}
\newcommand{\ee}{\end{equation}}
\newcommand{\beq}{\begin{eqnarray}}
\newcommand{\eeq}{\end{eqnarray}}

\def\nue{\mathrel{{\nu_e}}}
\def\numu{\mathrel{{\nu_\mu}}}
\def\nutau{\mathrel{{\nu_\tau}}}

\def\barnumu{\mathrel{{\bar \nu}_\mu}}

\def \gta {\mathrel{\vcenter{\hbox{$>$}\nointerlineskip\hbox{$\sim$}}}}

\def\t13{\mathrel{{\theta_{13}}}}
\def\y12{\mathrel{{\tan^2 \theta_{12}}}}
\def\c2{\mathrel{{\chi^2 }}}

\newcommand{\n}{neutrino}
\newcommand{\ns}{neutrinos}

\newcommand{\fb}{FB}
\newcommand{\ic}{IceCube}

\begin{document}


\title{Neutrino Events at Icecube and the Fermi Bubbles}

\author{Cecilia Lunardini}
 \email{Cecilia.Lunardini@asu.edu}  
\affiliation{Department of Physics, Arizona State University, Tempe, AZ 85287-1504}

	\author{Soebur Razzaque}
	\email{srazzaque@uj.ac.za}
	\affiliation{Department of Physics, University of Johannesburg, PO Box
  524, Auckland Park 2006, South Africa}

\author{Kristopher T. Theodoseau}
 \email{Kristopher.Theodoseau@asu.edu}    
 \affiliation{Department of Physics, Arizona State University, Tempe, AZ 85287-1504}

\author{Lili Yang}
\email{lyang54@asu.edu}
\affiliation{Department of Physics, Arizona State University, Tempe, AZ 85287-1504}

\begin{abstract}
  We discuss the possibility that the IceCube neutrino telescope might
  be observing the Fermi Bubbles. If the bubbles discovered in gamma
  rays originate from accelerated protons, they should be strong
  emitters of high energy ($\gta $ GeV) neutrinos. These neutrinos are
  detectable as shower- or track-like events at a ${\rm Km^3}$
  neutrino observatory.  For a primary cosmic ray flux  with spectrum $\propto E^{-2.1}$ and  cutoff energy at or
  above 10 PeV, the Fermi Bubble flux substantially exceeds the
  atmospheric background, and could account for up to $\sim 4-5$ of
  the 28 events detected above $\sim $30 TeV at IceCube.  Running the
  detector for $\sim 5-7$ more years should be sufficient to discover
  this flux at high significance. For a primary cosmic ray flux with steeper spectrum, and/or lower cutoff
  energy, longer running times
  will be required to overcome the background.
\end{abstract}                            
 
\maketitle


Very recently, the study of the sky at high energy has received a new
impulse by the \ic\ observation of an excess of \n\ flux, relative to
the atmospheric \n\ background, above $\sim 30$ TeV
\cite{Aartsen:2013bka, Aartsen:2013pza}.  Of a total of 28 events, 21
are showers (or ``cascades''), mostly caused by electron and tau
neutrinos. For the remaining 7 events a muon track has been
identified, thus indicating a muon neutrino scattering.  Two of the
shower events exceed 1 PeV of deposited energy \cite{Aartsen:2013bka},
while the other 26 events are below $\sim 250$ TeV.  The 28 events
observed at \ic\ are a milestone in the field of \n\ astronomy, and
have triggered a feverish activity to understand their meaning and
their physics potential.

When comparing the data to theoretical models of high energy \n\
fluxes, it is natural to expect that multiple sources might contribute
to the observed signal. Although prompt atmospheric neutrinos could
fit some of the data \cite{Lipari:2013taa}, distant astrophysical
sources are the most natural explanation.  Cosmological emitters would
likely produce a uniform, diffuse flux, and the spatial distribution
of the events is compatible with this hypothesis.  Recent literature
discusses the cases of gamma ray bursts \cite{Razzaque:2013dsa} and
their lower-powered counterparts \cite{Razzaque:2003uv,
  Murase:2013ffa}, starburst galaxies\cite{Loeb:2006tw, Stecker:2006vz, Liu:2013wia}, cores of
active galactic nuclei \cite{Stecker:2013fxa,Winter:2013cla} and
active galaxies \cite{Kalashev:2013vba}, as well as intergalactic
shocks \cite{Murase:2013rfa}.

In addition to a diffuse extragalactic component, Galactic sources
would appear as anisotropies, spatially correlated with the Galactic
disk and bulge.  Recent analyses suggested spatial correlation of the
\ic\ data with unidentified TeV Galactic sources \cite{Fox:2013oza},
with the Galactic Center \cite{Razzaque:2013uoa} and the Fermi Bubbles
\cite{Razzaque:2013uoa, Ahlers:2013xia}.  Origination from known
Galactic TeV sources \cite{Gonzalez-Garcia:2013iha}, and from the
galactic plane in general \cite{Anchordoqui:2013qsi,Joshi:2013aua} has
also been studied.  Beyond the standard model, ideas include the decay
of heavy relics (Galactic and extragalactic)
\cite{Feldstein:2013kka,Esmaili:2013gha} and new physics contributions
to the neutrino cross sections \cite{Barger:2013pla}.

\begin{figure}[htbp]
\includegraphics[width=0.4\textwidth]{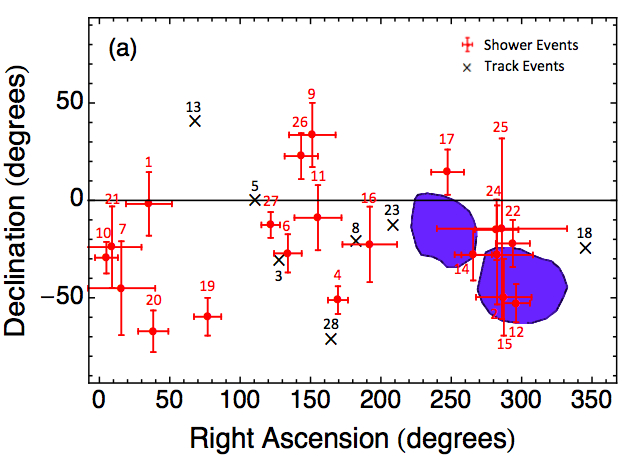}
\hbox{\hspace{5ex}\includegraphics[width=0.4\textwidth]{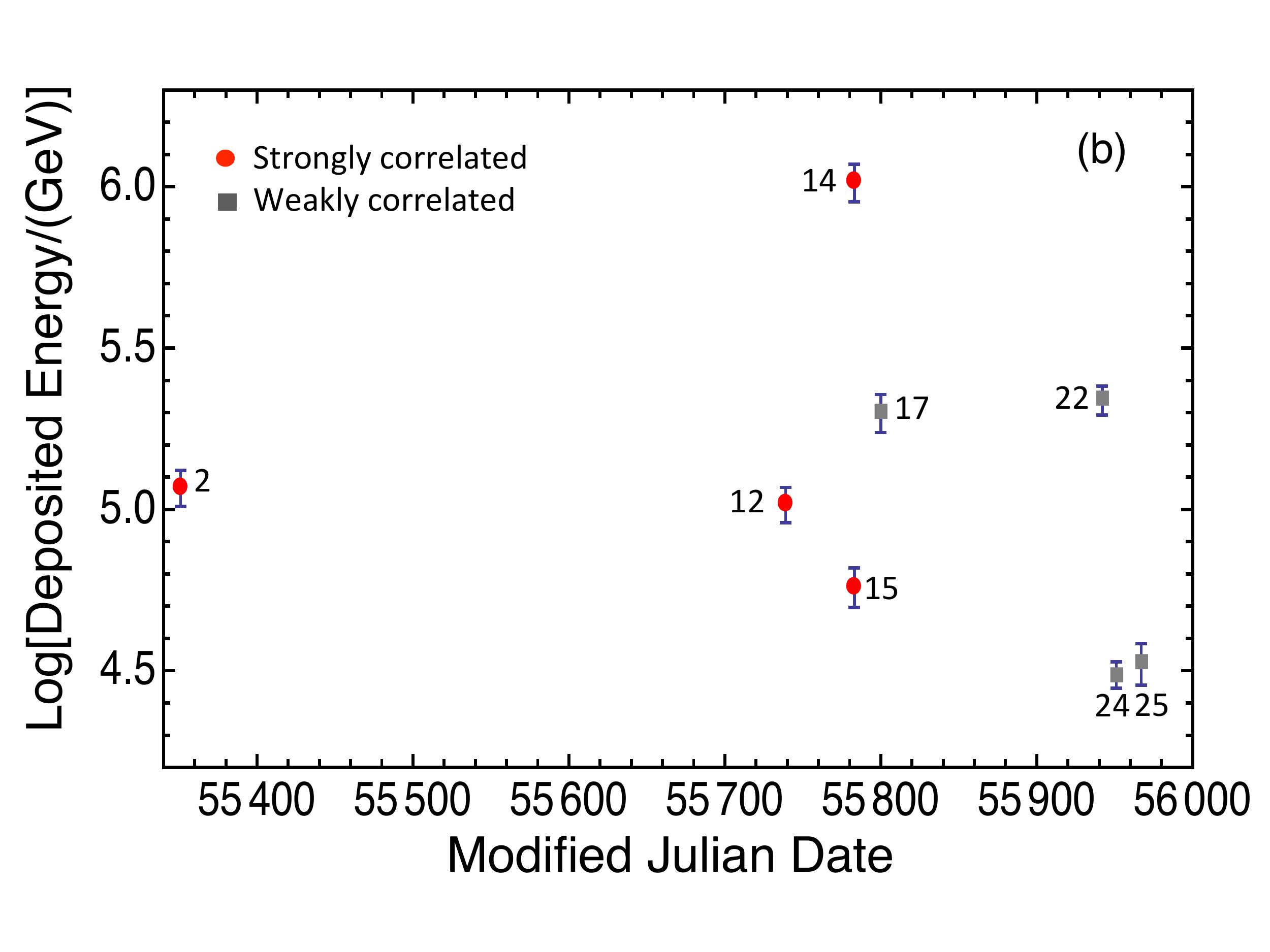}}

\caption{(a): The \ic\ events in equatorial coordinates, with their
  median angular errors, from \cite{Aartsen:2013pza}. The contours of
  the Fermi Bubbles are shown as well.  (b): The time and (deposited)
  energy distribution of the events that are spatially correlated with
  the bubbles. 
}
\label{icdata}
\end{figure}

\begin{figure}[htbp]
\centering
\includegraphics[width=0.36\textwidth]{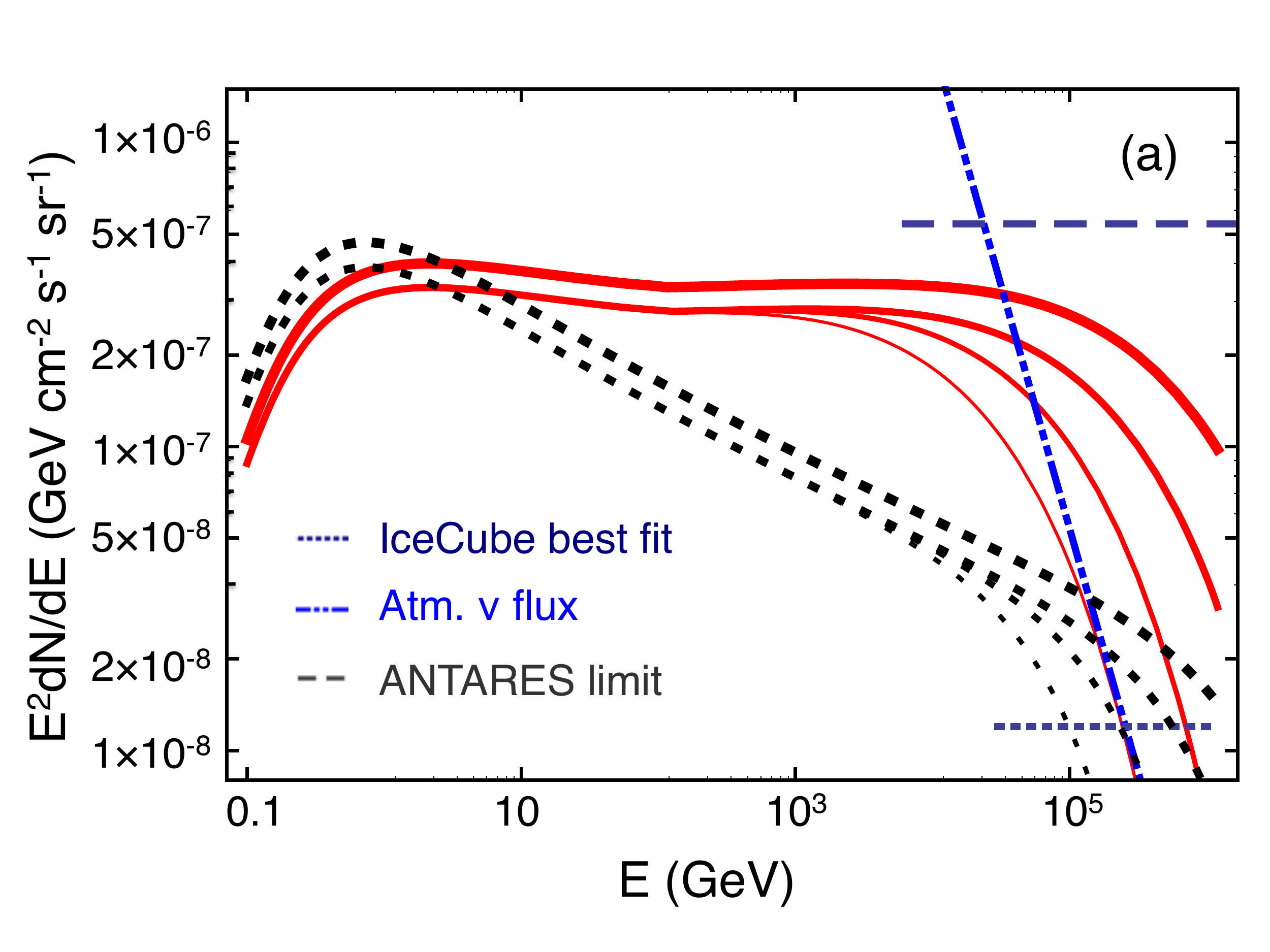}
\includegraphics[width=0.36\textwidth]{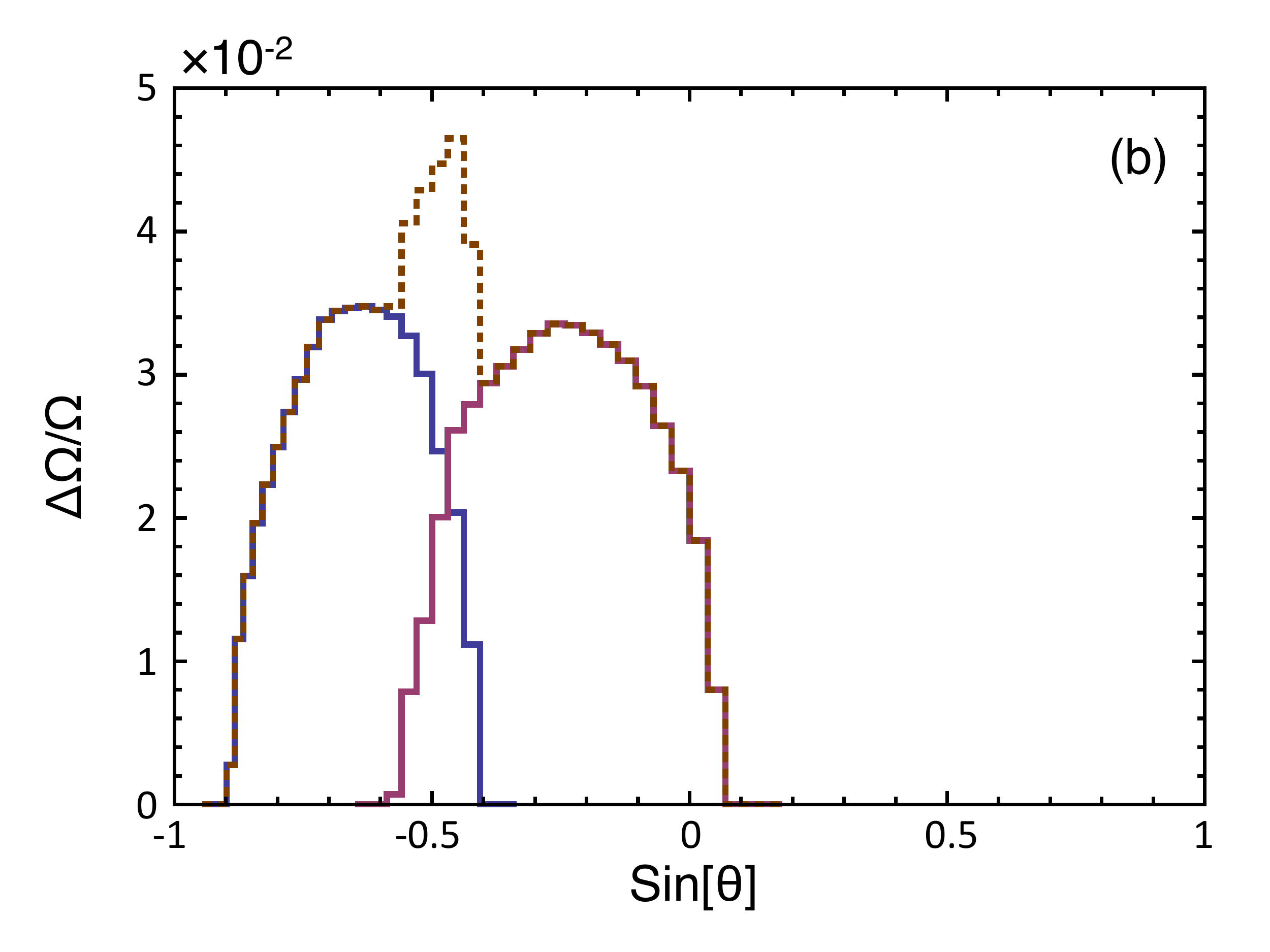}
\includegraphics[width=0.36\textwidth]{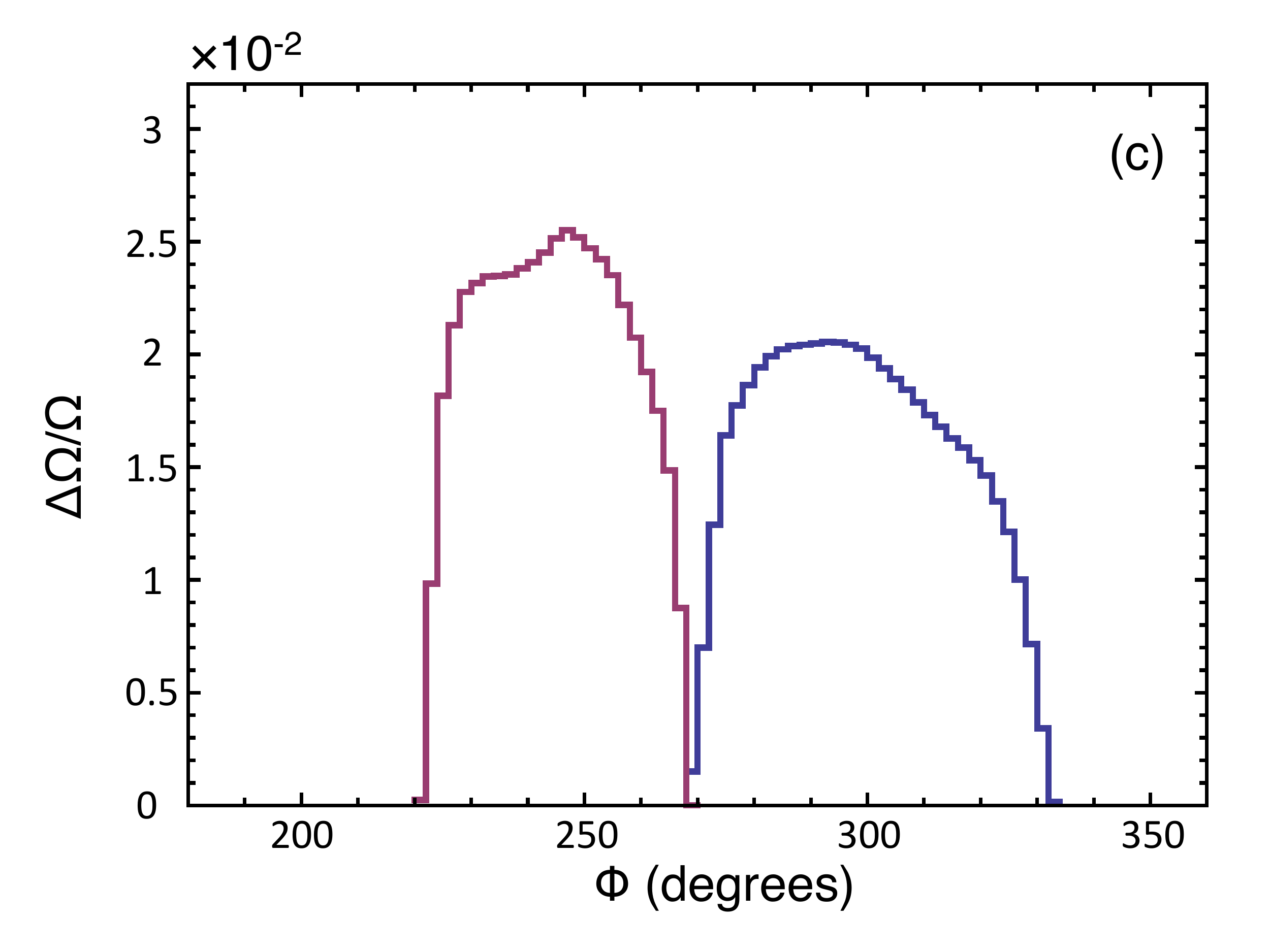}
\caption{ (a): The expected $\numu + \barnumu$ flux (solid lines) from
  the \fb\ (before oscillations), normalized to the gamma-ray flux, as
  a function of the energy, for different proton spectral
    indices, $k$, and different cutoffs of the primary proton
    spectrum, $E_0$.  Solid, red: $k=2.1$; dotted, black: $k=2.3$. For
    each we show, from thin to thick: $E_0=1,3,10,30$ PeV.  For
  comparison, we also show: (i) the atmospheric \n\ flux
  \cite{Honda:2006qj} averaged over $25^\circ$-$95^\circ$ zenith
  angle, (ii) the ANTARES upper limit \cite{Adrian-Martinez:2013xda}
  and (iii) the diffuse flux that best fits the \ic\ data \cite{Aartsen:2013pza}. The other two panels show
  the distribution (normalized to 1) of the flux in $\sin \theta$
  (with $\theta$ the declination angle) (b), and in the right
  ascension, $\phi$, (c), for each bubble (solid) and the total for
  both (dotted).  }
\label{fluxfig}
\end{figure}

The focus of this paper is to explore the detectability of the Fermi
Bubbles (\fb) at \ic.  Discovered in 2009 by Fermi-LAT
\cite{Su:2010qj}, the bubbles are extended gamma-ray sources of
globular shape, protruding symmetrically out of the Galactic Center
(GC) up to a distance of $\sim 9$ kpc.  Their origin, and the
production mechanism of gamma rays, are yet unknown.  Leaving aside
possible new physics
\cite{Cholis:2012fr,Okada:2013bna,Hooper:2013rwa,Huang:2013pda,Huang:2013apa},
concentrated high rate of supernova activity near the GC
\cite{Crocker:2010dg, Fujita:2013jda} or accretion of gas by the GC
black hole at a high rate in recent past \cite{Su:2010qj} are the two
main scenarios for bubble formation.  The observed gamma rays are
created either due to Compton scattering by highly-relativistic
electrons or due to decays of neutral pions created by interactions of
energetic baryons.  In the baryonic hypothesis, the gamma ray flux
from the bubbles should have a neutrino counterpart of similar
magnitude \cite{Crocker:2010dg, Lunardini:2011br}, that should be
detectable in muon tracks at a Km$^3$ detector in the northern
hemisphere \cite{Lunardini:2011br}.  Dedicated experimental work on
this is in progress
\cite{Adrian-Martinez:2013xda,Adrian-Martinez:2012qpa}, and an upper
limit has been placed by the Antares collaboration (see Fig.
\ref{fluxfig}) \cite{Adrian-Martinez:2013xda}.

For a Southern hemisphere detector like \ic, instead, the main
signature of the bubbles should be showers, thanks to the reduced
background and increased shower effective area of the detector for
down-going neutrinos compared to tracks \cite{Aartsen:2013bka,
  Aartsen:2013pza}. Here we present the first quantitative study of
the shower as well as down-going track events expected from the Fermi
Bubbles, both as a possible interpretation of some of the \ic\ data,
and as prediction for future searches with enhanced detector
configuration and exposure.

Seen from Earth, the Fermi Bubbles appear as extended sources in the
Southern sky (Fig.~\ref{icdata}) subtending a total solid angle
$\Omega_{FB}\simeq 0.808~{\rm sr}$ \cite{Su:2010qj}.  Interestingly,
their gamma ray emission per unit solid angle is roughly uniform over
the extent of the bubbles \cite{Su:2010qj}, and the same feature is
expected for the \n\ emission as well \cite{Lunardini:2011br}.

To estimate a possible correlation between the \ic\ events and the
\fb, we compare the bubbles coordinates with the reconstructed
coordinates of the \ic\ events and their median angular errors
\cite{Aartsen:2013pza}, see Fig.~\ref{icdata}.  It appears that
$N_{s}=4$ events (events number 2, 12, 14, 15) have their central
position value inside the bubbles (``strongly correlated'', meaning
higher likelihood of originating from the \fb), and $N_w=4$ (number
17, 22, 24, 25) are compatible with the bubbles within the error
(``weakly correlated'', or lower likelihood). Therefore $N=8$ is a
conservative upper limit for the number of events from the \fb, to be
compared with theoretical predictions.  Note that one of the strongly
correlated events, event number 14, has $\sim$1 PeV of deposited
energy \cite{Aartsen:2013pza}.
 
To calculate the event rate in \ic\ due to the \fb, we use the
neutrino fluxes from Ref.~\cite{Lunardini:2011br}.  These fluxes are
derived from fitting the gamma-ray data using $pp$ interactions of
cosmic-ray protons in the bubbles with the ambient gas.  A proton
  spectrum $dN/dE\propto E^{-k}$ was used, with a cutoff energy $E_0$,
  motivated by the maximum energy to which supernova remnants can
  accelerate cosmic ray protons.  Theoretical estimates of $E_0$ vary
  from 1 PeV, at the ``knee'' of the cosmic-ray spectrum, to 100 PeV
  \cite{Ptuskin:2010zn}.  The hard $\gamma$-ray spectrum of the \fb\ is
  best represented with $k=2.1$, which is also favored by
  shock-acceleration theories.  This is our default flux model unless
  otherwise specified.  Given rather limited range of $\gamma$-ray
  data, a steeper $k=2.3$ proton spectrum is also compatible with
  observation. As shown in Fig.~\ref{fluxfig} (a), the fluxes differ
significantly above $\sim 200$ GeV (above the range of gamma-ray data)
depending on $E_0$.  Fig.~\ref{fluxfig} (a) also shows our most
optimistic flux model (solid curve), obtained with $E_0 = 30$ PeV, and
a $\sim 20\%$ increase of the normalization of the whole flux, which
is allowed by the uncertainty in the gamma ray data. All results
quoted for $E_0=30$ PeV will refer to this model. For comparison, in Fig. ~\ref{fluxfig} (a) we show the diffuse flux (at the detector after oscillation) that best fits the IceCube data \cite{Aartsen:2013pza}. Note that this flux refers to fitting the entire data sample in the assumption of a diffuse, uniform flux over the whole sky. It would be interesting to fit the data that are spatially correlated with the \fb\ to find the level of flux required to reproduce them. At present, however, this can not be done in the absence of more detailed information on the \ic\ effective area and exposure. 

The initial (pre-oscillation) flavor composition of the flux is $\nue$
: $\numu$ : $\nutau$ $= \epsilon : 1 : 0$, with $\epsilon$ increasing
from $\epsilon \simeq 0.57$ at $E=1$ TeV to $\epsilon \simeq 0.88$ at
$E=1$ PeV. This is explained by how energy is shared between the
products of pion decay at different energies \cite{Lipari:1993hd}.
After oscillations (averaged vacuum oscillations, matter effects are
negligible) the flavor ratios are close to $\nue$ : $\numu$ : $\nutau$
=$ 1 : 1 : 1$, with deviations up to $\sim 30\%$ at $E \sim 1$ PeV.

Because the emission is uniform over the \fb\ surfaces, the fraction
of flux in a solid angle bin, $\Delta F/F$ is given by the fraction of
solid angle, $\Delta \Omega/\Omega_{FB}$.  This is shown in Fig.
\ref{fluxfig}(b) and (c).

\begin{figure}[htbp]
\centering
\includegraphics[width=0.4\textwidth]{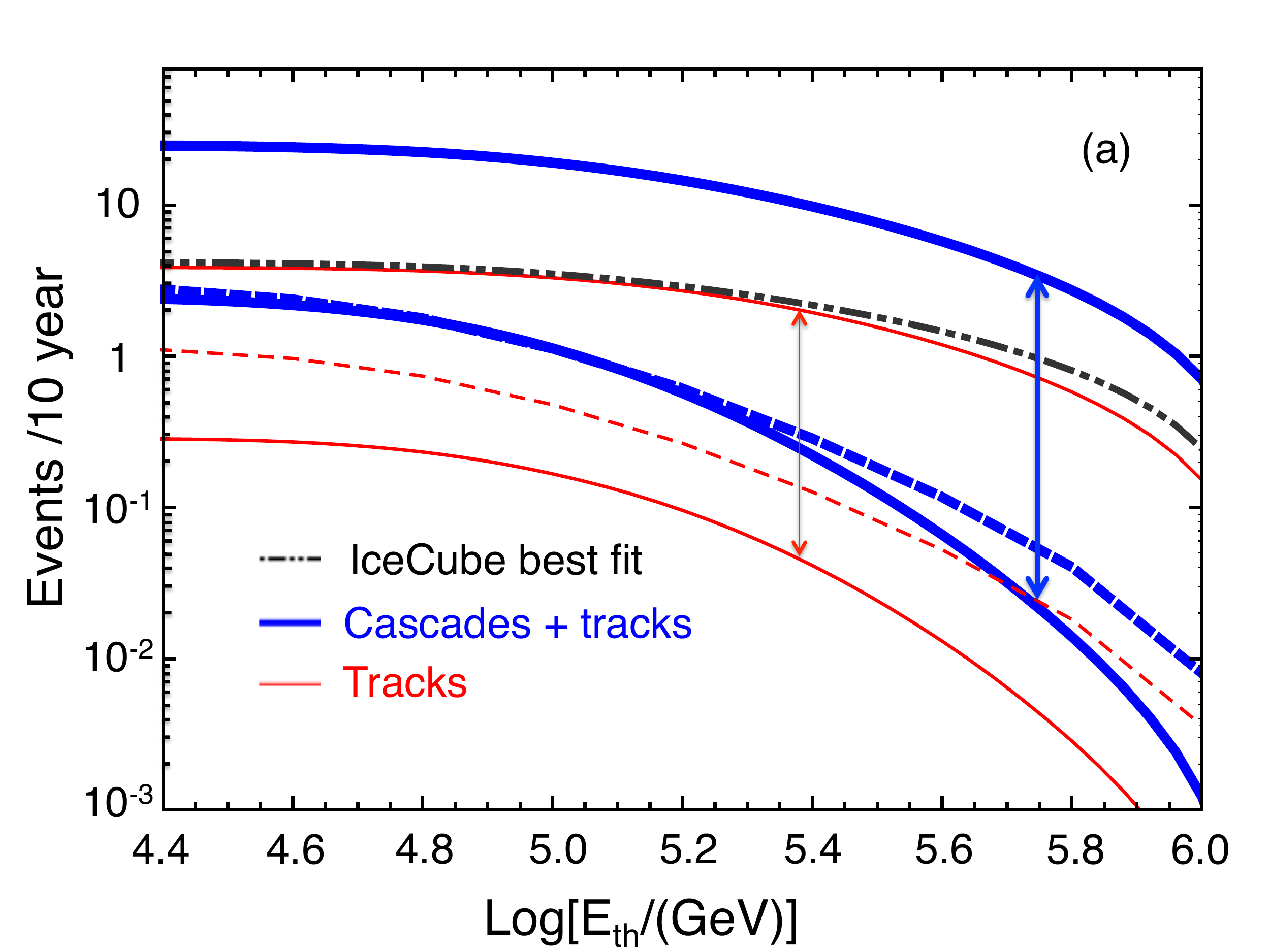}
\includegraphics[width=0.4\textwidth]{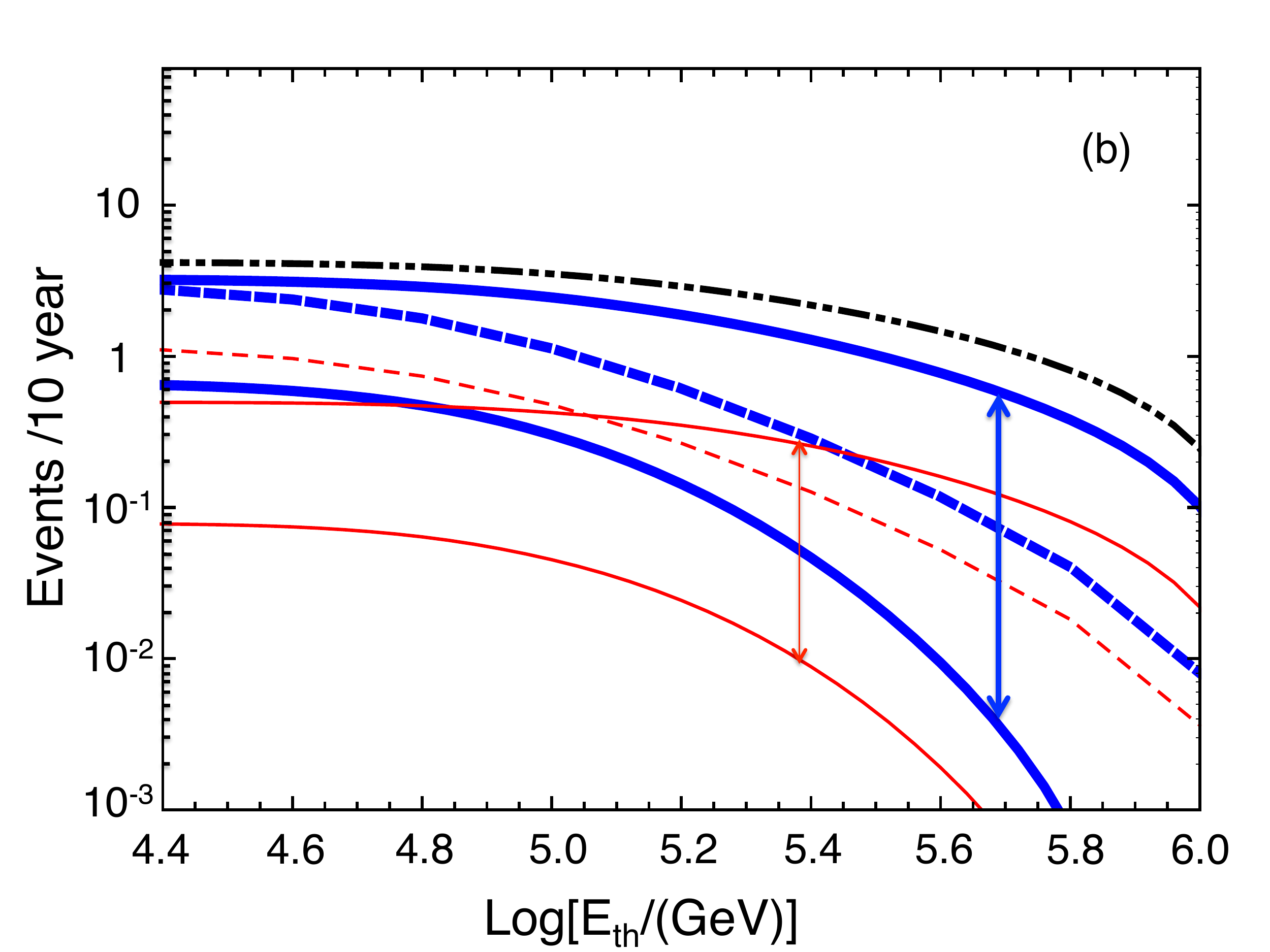}
 \caption{Events expected at \ic\ per decade,
   as a function of the \n\ energy threshold $E_{th}$, for the
   primary proton spectrum index $k=2.1$ (upper panel) and
   $k=2.3$ (lower panel).  Solid: \fb\ signal, for the total of
   shower- and track-like events (thick) and for track-like
   events only (thin).  The arrows indicate the effect of varying
   the primary spectrum cutoff in the interval $E_0 = 1 - 30$
   PeV.  Dashed: the same but from atmospheric fluxes.
   Dot-dot-dashed: showers- and track-like events from the \ic\
   best-fit flux in Fig. \ref{fluxfig}(a). }
\label{ratefig}
\end{figure}

\begin{figure}[htbp]
\centering
\includegraphics[width=0.4\textwidth]{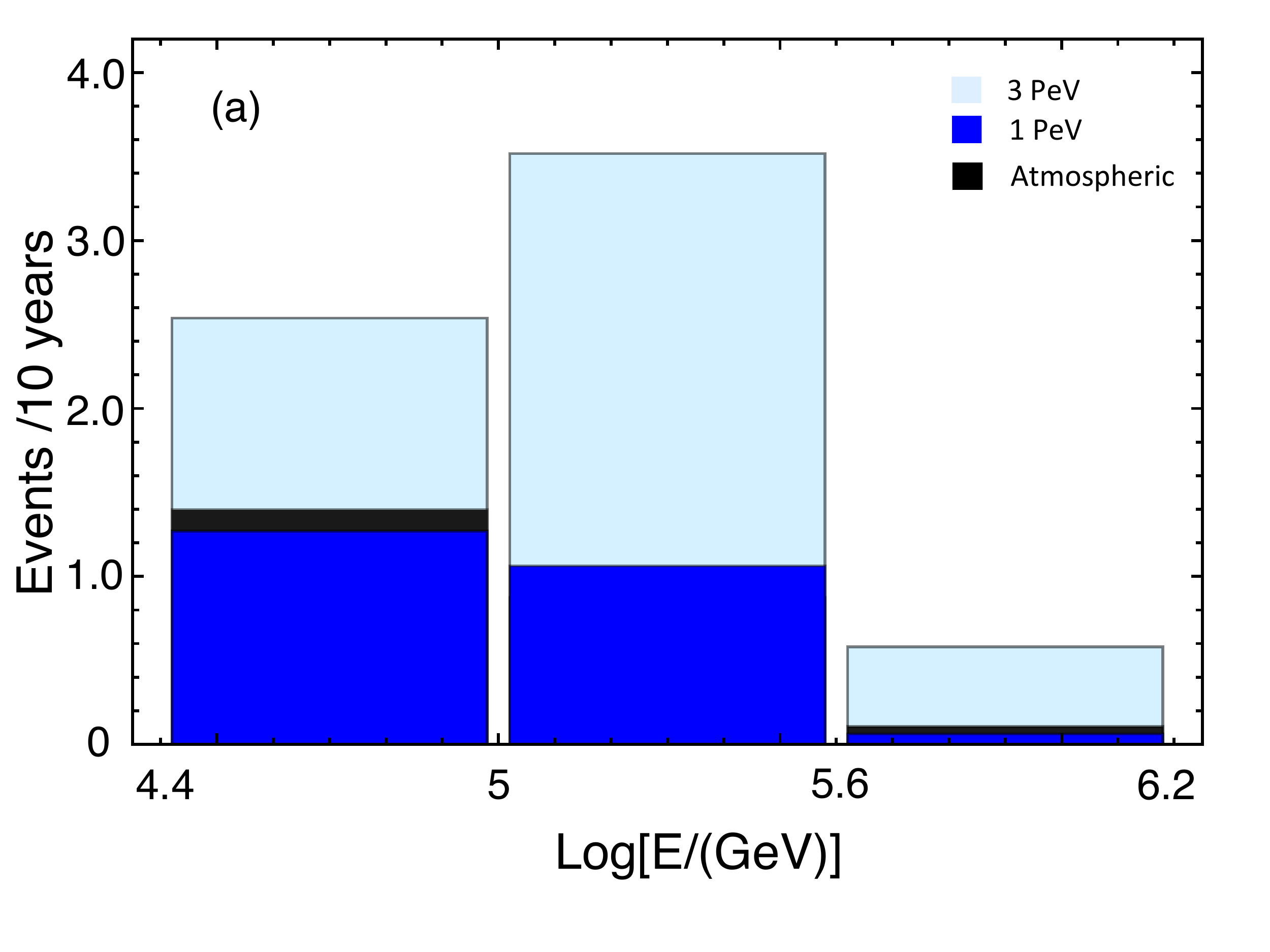}
\includegraphics[width=0.4\textwidth]{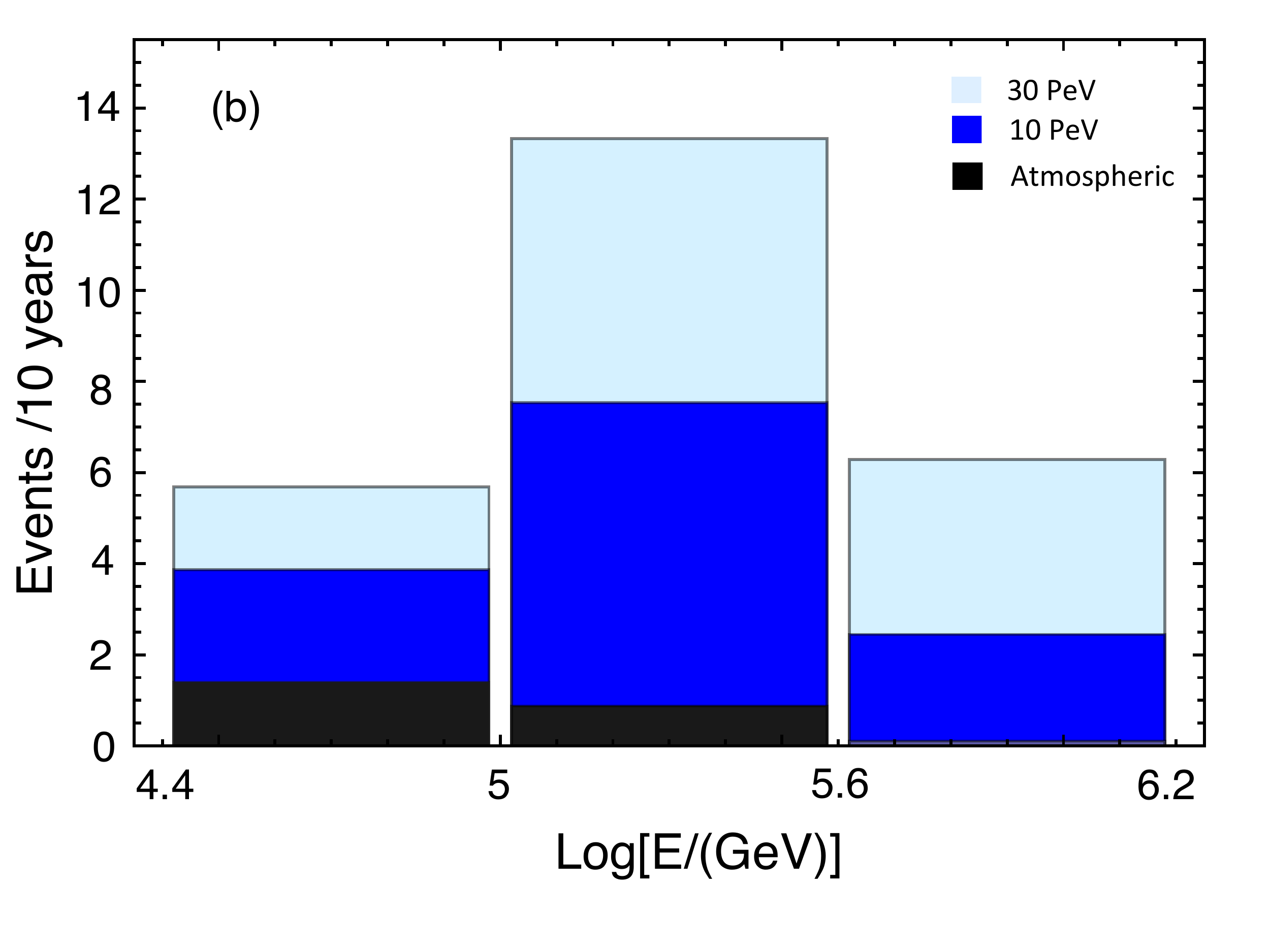}
\caption{Expected neutrino energy distribution of the total shower- and
  track-like events per decade, for signal (with $k=2.1$)  and for atmospheric background.  (a): $E_{0}=1, 3$ PeV ; (b): $E_{0}=10, 30$ PeV.  For $E_{0}=1$ PeV, signal and
  background are very close in all bins.   }
\label{histofig}
\end{figure}

To establish the significance of the \fb\ signal, one should consider the main backgrounds, i.e., atmospheric muons and atmospheric neutrinos.  For the former, the background level depends on the detector veto, and could change with future technological advances. We refer to \cite{Aartsen:2013pza} for this.  Here we model the atmospheric neutrino background 
using the
neutrino flux prediction by Honda et
al. \cite{Honda:2006qj} (which is a good fit of \ic's atmospheric data
\cite{Abbasi:2010ie}), extrapolated at high energy, and a $\numu/\nue$
ratio of about 14 \cite{Sinegovskaya:2013wgm}. We also consider the
flux to be symmetric in $\cos \theta_z$
\cite{Athar:2012it}. Oscillations are negligible at the energies and zenith
angles of interest \cite{Gaisser:1997eu}, therefore the $\nutau$
atmospheric flux is neglected altogether.  To account for the error on
the direction of arrival of the \ns,
we calculate the rate of atmospheric shower events over a solid angle
larger than $\Omega_{FB}$, obtained by encasing each bubble in a
rectangle in the $\theta$ and $\phi$ coodinates, and then enlarge such
rectangle by $\omega = 15^\circ$ (motivated by the detector's angular
resolution \cite{Aartsen:2013pza}) on each side.  The total solid
angle obtained in this way is $\Omega_{bckg}\simeq 2.75~ {\rm sr}$.  For track events, where the angular resolution is
less than a degree \cite{Aartsen:2013pza}, the angle $\Omega_{FB}$ is
used.  We find that shower- and track-like events contribute
comparably to the total background rate, because the predominance of
the $\numu$ species in the atmospheric flux compensates for the
smaller effective area for tracks.

Fig.~\ref{ratefig}(a) shows the expected number of signal and
  background events for $k=2.1$ and 10 years running time, above an
energy threshold $E_{th}$, as a function of $E_{th}$.  
We observe that, for $E_0 \gta  10~{\rm PeV}$, the signal rises above the background, by
up to $\sim$2 orders of magnitude for the most optimistic flux model.
Specifically, for $E_0 = 30~{\rm PeV}$ and $E_{th}= 10^{4.6}~{\rm
  GeV}$, we find 23 signal and 2 background events, amounting to a
$\sim 4.4~\sigma$ excess due to the FB.  For the same parameters, a
significance of $3\sigma$ would be obtained with about 7 years of
running time.  The time needed for discovery might be shorter with the
use of detailed statistical analyses of the spatial correlation with
the bubbles, and/or if a compatible excess is observed in track events
at a detector in the Northern hemisphere
\cite{Lunardini:2011br,Adrian-Martinez:2012qpa}. 
For the most conservative spectrum, $E_0 = 1~{\rm PeV}$, the
background is comparable to the signal for all thresholds, therefore, detections
prospects are poor.  For the steeper spectrum, $k=2.3$ (fig. ~\ref{ratefig}(b)), conclusions are similar, overall.  However, even for the most optimistic spectrum, the signal/background ratio is  modest, and becomes significant only above $\sim 10^5$~GeV, where the event rate is small.  

Fig. \ref{histofig} gives the distribution of signal and background events per decade in bins of neutrino energy \footnote{The observed energy could be lower by a factor of 3-4 than the neutrino energy for neutral current events \cite{Aartsen:2013vja}. Therefore, our fig. \ref{histofig} is not representative of the actual observed data spectrum. However, it should be a reasonable approximation, considering our choice of very wide energy bins. }. The width of the bins are chosen such that in each bin the highest energy is 4 times the lowest energy, which is roughly the maximum uncertainty in reconstructing the neutrino energy from the deposited energy in case of neutral current interactions \cite{Aartsen:2013vja}. Overall, fig. 4 confirms the results of fig. 3(a); it also shows that most of the events are expected the bin $\log(E/{\rm GeV}) = 5-5.6$, due to a sharp rise of the effective area below $\sim 1~{\rm PeV}$ \cite{Aartsen:2013pza} and an $E_0$-dependent exponential drop of the flux at high energies.

\begin{figure}[htbp]
\centering
\includegraphics[width=0.4\textwidth]{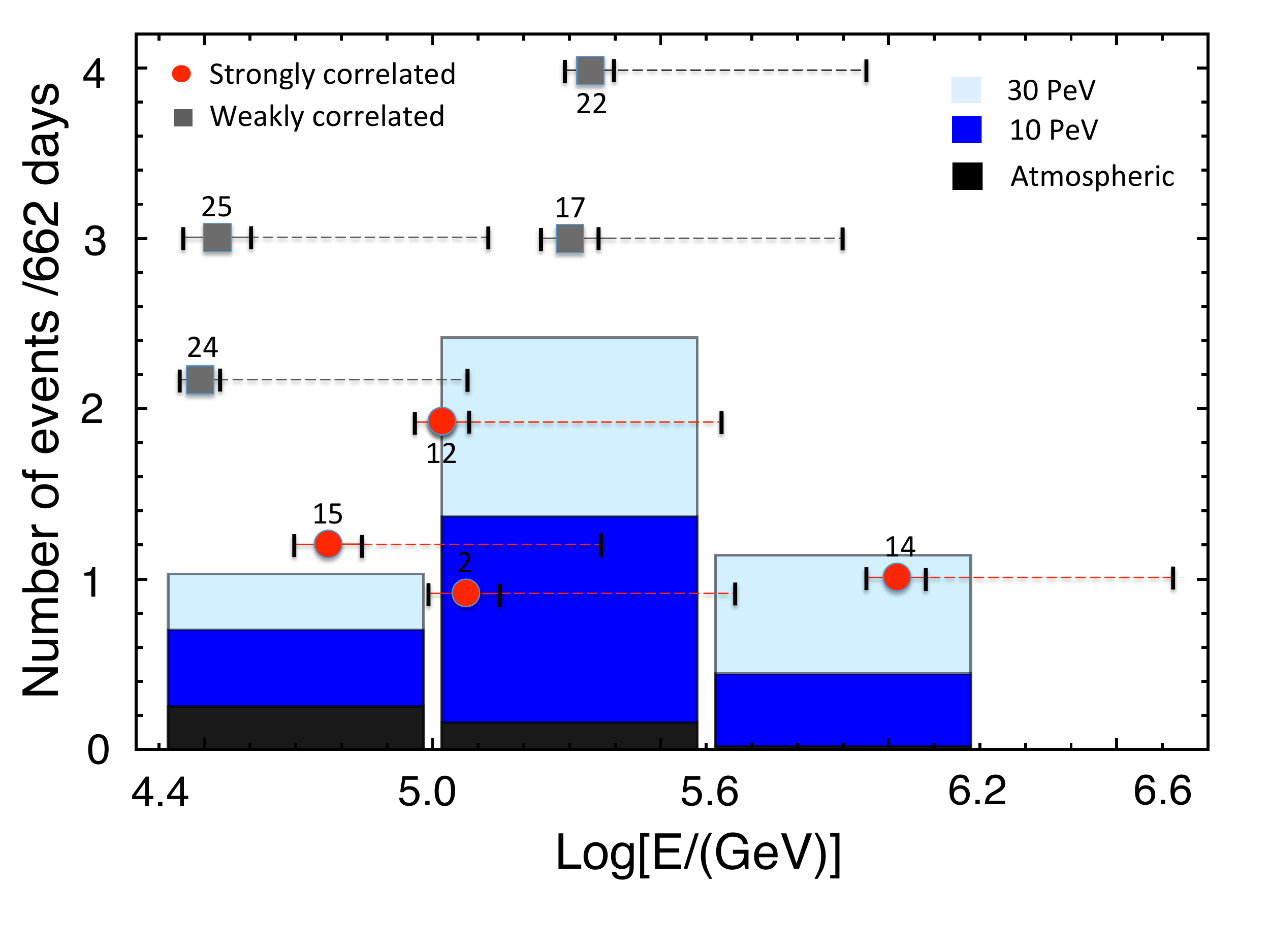}
\caption{The same as fig. \ref{histofig}(b) for the \ic\ running time of 662 days. The \ic\ events that correlate with the \fb\ (from Fig.~\ref{icdata}) are shown for comparison.   Their coordinates on the vertical axis give a visual representation of the number of events for which the central value of the observed energy  falls in a given bin.  The solid (dashed) error bars represent, respectively, the error on the observed energy and the factor of $\sim 3-4$ difference between \n\ energy and observed energy for neutral current events.  }
\label{ice3rate}
\end{figure}

Let us now apply our results to the \ic\ data, from the recent
662 days search \cite{Aartsen:2013pza}.  Table
\ref{tabrate}and fig. \ref{ice3rate} show the expected number of events for signal and
background.  For the total of shower- and track-like events, less than
one atmospheric background event is expected. The \fb\ signal rises above one
event for $E_0>3$ PeV, and for $E_0\geq 10$ PeV, it starts to be close
to the measured rate.  In particular, for $E_0=30$ PeV,  we expect
$N\sim 3$ and $N\sim 1$ events below and above $E=10^{5.6}~{\rm GeV} \simeq 400~{\rm TeV}$ of
neutrino energy respectively.  This is intriguingly close, in number and energy distribution, to the
observation of the $N_s=4$ events strongly correlated with the \fb\ (fig. \ref{ice3rate}).

\begin{table}[htdp]
\caption{Showers + track-like events expected in three bins of \n\ energy, from the atmospheric background and from the \fb\ (for different primary spectrum cutoff, $E_0$) for the 662 days \ic\ search. The numbers in brackets refer to track-like events only.  }
\begin{center}
\begin{tabular}{|c|c|c|c|c|}
\hline 
  & \multicolumn{3}{|c|}{$\log(E/{\rm GeV})$} & total \\ 
 \hline
 & $4.4 -5 $ & $5 - 5.6 $  & $5.6 - 6.2 $ &  \\
\hline
\hline
$E_0/{\rm PeV} = 1$ &  0.23 & 0.19 & 0.01 & 0.43 \\
                         &   [0.02] & [0.03] & [0] & [0.05] \\
 \hline
$E_0/{\rm PeV}  = 3$  &  0.46 & 0.64 & 0.11 & 1.2 \\
                         &   [0.04] & [0.1] & [0.02] & [0.16 ]\\
  \hline
$E_0/{\rm PeV} = 10$  & 0.7 & 1.37 & 0.44 & 2.51 \\
                          &  [0.07] & [0.21] & [0.09] & [0.37] \\
                           \hline
$E_0/{\rm PeV} = 30$  &   1.03 & 2.42 & 1.14 & 4.59 \\
                            & [0.1] & [0.38] & [0.24] & [0.72] \\
                            \hline
 Background &  0.25 & 0.16 & 0.02 & 0.43 \\
                            &   [0.07] & [0.06] & [0.01] & [0.14] \\
\hline 
\hline 
\end{tabular}
\end{center}
\label{tabrate}
\end{table}%

In synthesis, has Icecube already detected the \fb?  The answer might
be yes, if the \n\ spectrum is relatively hard, coming from a primary
proton flux  that falls like $E^{-2.1}$ and has a cutoff above 10 PeV or so.  $E_0 \simeq 10-30$
  PeV seems to best fit the data, specially the events strongly
  correlated with the \fb.  Observation of a \n\ flux from the \fb\
  may provide clues to the maximum limit of particle acceleration in
  supernova remnants, which are thought to be the origin of energetic
  protons in the \fb\ and which are not widely discussed as sources of cosmic rays above 1 PeV.  Note, however, that protons (and in effect
  heavy nuclei, if present) in the \fb\ are thought to lose all their
  energy by $pp$ interactions over the life time of the bubbles
  (several Billion years) in the hadronic model \cite{Crocker:2010dg}.
  Thus \fb\ are not expected to contribute significantly to the
  observed cosmic-ray spectrum, which is dominated by heavy nuclei above the ``knee'' at $\sim 1$ PeV.

Our model predicts that up to $\sim$5 of the observed \ic\ events
might be due to the \fb.  Like other models with a strong Galactic
contribution, this implies that the extragalactic, diffuse, flux
required to explain rest of the data should be lower compared to the
case with no galactic flux.  Considering that about $\sim 10$ events
in \ic\ are likely to be background \cite{Aartsen:2013pza}, the
diffuse flux normalization would have to be smaller by $\sim 4/(28 -
10) = 22\%$.  This figure is insignificant with the current
statistics, but might be nevertheless important to consider when
looking in perspective for the future. 

With higher statistics, the \fb\ should clearly manifest themselves
with an excess of events correlated with their position and extent in
the sky (Fig. \ref{icdata}).  No other phenomenon would have such a
signature.  The statistics needed to have a significant detection of
the \fb\ depends on the level of the diffuse \n\ flux from other
sources (other than atmospheric background), however, at least for the
most optimistic scenario ($E_0=30$ PeV) 7-10 years time should be
sufficient, see fig. \ref{ratefig}.

The \fb\ signal will be strongly substantiated by a northern
hemisphere detector like the future Km3Net
\cite{Adrian-Martinez:2012qpa}, which will be at a nearly optimal
location to look for track-like events from the bubbles
\cite{Lunardini:2011br}.  More than 300 events per decade are expected
for $E_0=10$ PeV. The complementarity of \ic\ and Km3Net will help to
resolve a number of uncertainties and degeneracies (for example, the
track events at Km3Net will have a better angular resolution, thus
helping to separate the \fb\ from other possible Galactic sources).

If the \fb\ are confirmed to be strong \n\ emitters, the implications
on the physics of our galaxy would be profound. In particular, this
would support the idea of a long time scale of the activity of the
Galactic center, $\sim 10^9$ years. This is the time required to form
the \fb\ in the hadronic model \cite{Crocker:2010dg}, as opposed to
the much shorter time (millions of years) required in a leptonic model
of the bubbles.

We thank Albrecth Karle, Mariola Lesiak-Bzdak, Jakob van Santen and
Nathan Whitehorn.  C.L. and L.Y. acknowledge the
National Science Foundation grant number PHY-1205745.
K.T. acknowledges the ASU/NASA Space Grant  2013 for partial support. S.R. acknowledges support from the National Research Foundation (South Africa) grant CPRR 2014 number 87823.



\end{document}